\documentclass[fleqn,10pt]{wlscirep}
\usepackage{geometry}
\usepackage{etoolbox}
\usepackage{graphicx}
\usepackage{color}
\usepackage{graphicx}
\usepackage{amsmath}
\usepackage{capt-of}
\usepackage{caption}
\usepackage{slashbox}

\title{Coherent control of Optical limiting in atomic systems}

\author[1]{Mohsen Ghaderi Goran Abad}
\author[1]{Mahsa Mahdieh}
\author[1]{Mohadeseh Veisi}
\author[1]{Hamid Nadjari}
\author[1,*]{Mohammad Mahmoudi}
\affil[1]{Department of Physics, University of Zanjan, University Blvd., 45371-38791, Zanjan, Iran}
\affil[*]{mahmoudi@znu.ac.ir}

\begin{abstract}
Generation and control of the reverse saturable absorption (RSA) and optical limiting (OL) are investigated in a four-level Y-type quantum system. It is demonstrated that the applied laser fields induce the RSA and it can be coherently controlled by either intensity or frequency of the applied laser fields. The effect of the static magnetic field on the induced RSA is studied and we obtain that it has a constructive role in determining the intensity range in which the OL is established in the system. In addition, we find that the OL threshold can be decreased either by increasing the length of the medium or by getting the atomic system denser. Finally, Z-scan technique is presented to confirm our theoretical results. The proposed scheme can be used in designing the coherent optical limiters with controllable threshold and intensity range of OL.

\end{abstract}
\begin{document}
\flushbottom
\maketitle

\thispagestyle{fancy}

Atomic coherence offers a systematic basis for the fundamental concepts of the coherent control. Optical devices are generally based on manipulation of the phase, polarization, propagation direction and also intensity of the optical light which can be done by atomic coherence due to the applied laser fields. Intensity manipulation of the light is a major basis of designing all-optical switches \cite{Zhang,Tutt} and optical limiters \cite{Tutt,Sun}. Optical limiters have been receiving much attention for the increasing demand in protection of optical components in optically based devices. All optical sensors involved in the optical devices may be vulnerable for their sensitivity to the light intensity. In fact, human eye, sensors and other optical sensitive elements have intensity threshold above which laser-induced damage happens.  In optical limiters the transmission of the light reduces or even becomes constant for the input intensities higher than the threshold intensity. Thus, the presence of optical limiters to restrict the intensity of the incident laser beam is completely requisite prior to the sensors and direct viewing devices. Optical limiters protect the sensors from the damages due to the higher intensity laser pulse by extension their intensity range to operate under rougher conditions.

Various techniques, i.e., two photon absorption \cite{Li,Ehrlich,Rao}, nonlinear scattering \cite{Venkatram,Kumar} have been reported for generating the OL.  The basic mechanism to establish the OL is the RSA \cite{Kobyakov,Dong,Lepkowicz,Azzam} in which, unlike the saturable absorption (SA), the absorption increases by increasing the incident intensity. In general, the RSA can occur when the absorption of the excited state is large compared to the absorption of the ground state. On the contrary, SA is the dominant phenomenon in the system. Note that in SA materials, the absorption reduces by increasing the intensity due to depletion of the ground state, leading the materials to be more transparent. The RSA and OL have been observed in numerous compounds such as organic materials \cite{Sun,Tong}, $C_{60}$ solution \cite{Mishra}, nano materials \cite{Pan,Wang,Chen,Muller} and semiconductors \cite{Tintu}.  In addition, most studies on the OL have been carried out considering molecular systems with the aid of the rate equation approach \cite{Kiran,Lu,Allam,Aziz}.

In this paper, unlike the previous reported works, we introduce a coherently controllable optical limiter using the atomic systems in a four-level Y-type configuration. It is demonstrated that the RSA is coherently induced by the laser fields and the conditions are provided for preparing the OL in the system. It is shown that all characteristics of the induced OL such as the intensity range and the threshold intensity can be controlled by either intensity or frequency of the laser fields. Moreover, the effect of the static magnetic field on the OL is studied and it is illustrated that the RSA and the OL regions are extended by increasing the magnitude of the static magnetic field. In addition, we show that the OL threshold decreases  either by increasing the length of the medium or by getting the atomic medium denser. Finally, Z-scan technique is presented to confirm our theoretical results.

\section*{Model and Equations}
\begin{figure}[htbp]
\centering
  \includegraphics[width=.4\textwidth]{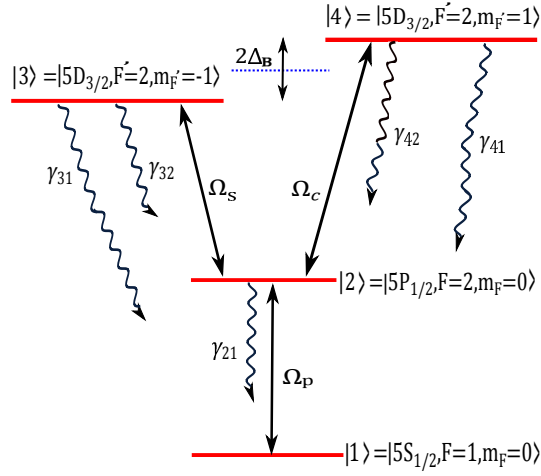}
  \caption{\small A schematic of a four-level Y-type quantum system driven by a weak probe field and two coupling fields with Rabi frequencies $\Omega_p$, $\Omega_s$ and $\Omega_c$ respectively.}\label{f1}
\end{figure}

The proposed realistic atomic system is a four-level Y-type quantum system which can be established in $5S_{1/2}$,  $5P_{1/2}$ and  $5D_{3/2}$ lines of $^{87}$Rb atoms as shown in Fig. \ref{f1}. Two lower states are $|1\rangle=|5S_{1/2},(F=1,m_{F}=0)\rangle$ and $|2\rangle=|5P_{1/2},(F=2,m_{F}=0)\rangle$, separated by 377 THz. Two degenerate states $|3\rangle=|5D_{3/2}, (F'=2,m_{F'}=-1)\rangle$ and $|4\rangle=|5D_{3/2}, (F'=2,m_{F'}=1)\rangle$ are chosen as the excited states. Here, $F$ and $F'$ are the quantum numbers of the total angular momentum and $m_{F(F')}$ denotes magnetic quantum number of the corresponding states. A weak linearly probe field, $\vec{E_p}(z,t)=\vec{\varepsilon}_p(z) e^{-i(\omega_{p}t-k_{p}z)}+c.c.,$ with wave vector $k_{p}$ and polarization in $\hat{x}$ direction, drives the transition $|1\rangle\leftrightarrow|2\rangle$ with the Rabi frequency $\Omega_{p}=\vec{\mu}_{21}. \vec{\varepsilon_p}/\hbar$. The transition $|2\rangle\leftrightarrow|3\rangle$ ($|2\rangle\leftrightarrow|4\rangle$) is coupled by the strong left (right) circularly polarized coupling field with the Rabi frequency $\Omega_{s}=\vec{\mu}_{32}.\hat{\epsilon}_{-}E_{s}/\hbar$ ($\Omega_{c}=\vec{\mu}_{42}.\hat{\epsilon}_{+}E_{c}/\hbar$). $\varepsilon_p$ and $E_{i}(i=s, c)$ are the amplitude of the probe and coupling fields, respectively.  $\epsilon_{\pm}$ stands for the polarization unit vectors of the coupling fields. A static magnetic field is also employed to remove the degeneracy of the states $|3\rangle$ and $|4\rangle$ by 2$\hbar\Delta_B=2m_{F}g_{s}\mu_{B}B$  where $\mu_{B}$ is Bohr magneton and $g_{s}$ is Land$\acute{e}$ factor.

The Hamiltonian of the considered system in the dipole and rotating wave approximations can be written as
\begin{equation}\label{e1}
    V_{I}=-\hbar(\Omega^{\ast}_{p}e^{-i\Delta_{p}t}|2\rangle\langle1|+\Omega^{\ast}_{s}e^{-i(\Delta_{s}+\Delta_{B})t}|3\rangle\langle2|+\Omega^{\ast}_{c}e^{-i(\Delta_{c}-\Delta_{B})t}|4\rangle\langle2|)+h.c.,
\end{equation}
where $\Delta_{p}=\omega_p-\omega_{21}$, $\Delta_{s}=\omega_s-\omega_{32}$ and $\Delta_{c}=\omega_c-\omega_{42}$ are the detunings of the applied fields frequencies and the central frequencies of the corresponding transitions.
The density matrix equations of motion can be written as follows
\begin{eqnarray}\label{e2}
\dot{\rho}_{11}&=&i\Omega_{p}\rho_{21}-i\Omega^{\ast}_{p}\rho_{12}+\gamma_{21}\rho_{22}+\gamma_{31}\rho_{33}+\gamma_{41}\rho_{44},\nonumber\\
\dot{\rho}_{33}&=&i\Omega^{\ast}_{s}\rho_{23}-i\Omega_{s}\rho_{32}-\Gamma_{1}\rho_{33},\nonumber\\
\dot{\rho}_{44}&=&i\Omega^{\ast}_{c}\rho_{24}-i\Omega_{c}\rho_{42}-\Gamma_{2}\rho_{44},\nonumber\\
\dot{\rho}_{21}&=&i\Omega^{\ast}_{p}(\rho_{11}-\rho_{22})+i\Omega_{s}\rho_{31}+i\Omega_{c}\rho_{41}-[\frac{\gamma_{21}}{2}-i\Delta_{p}]\rho_{21},\nonumber\\
\dot{\rho}_{31}&=&i\Omega^{\ast}_{s}\rho_{21}-i\Omega^{\ast}_{p}\rho_{32}-[\frac{\Gamma_{1}}{2}-i(\Delta_{s}+\Delta_ B+\Delta_{p})]\rho_{31},\nonumber\\
\dot{\rho}_{32}&=&i\Omega^{\ast}_{s}(\rho_{22}-\rho_{33})-i\Omega_{p}\rho_{31}-i\Omega^{\ast}_{c}\rho_{34}-[\frac{\Gamma_{1}+\gamma_{21}}{2}-i(\Delta_{s}+\Delta_ B)]\rho_{32},\nonumber\\
\dot{\rho}_{41}&=&i\Omega^{\ast}_{c}\rho_{21}-i\Omega^{\ast}_{p}\rho_{42}-[\frac{\Gamma_{2}}{2}-i(\Delta_{c}+\Delta_{p}-\Delta_ B)]\rho_{41},\nonumber\\
\dot{\rho}_{42}&=&i\Omega^{\ast}_{c}(\rho_{22}-\rho_{44})-i\Omega_{p}\rho_{41}-i\Omega^{\ast}_{s}\rho_{43}-[\frac{\Gamma_{2}+\gamma_{21}}{2}-i(\Delta_{c}-\Delta_ B)]\rho_{42},\nonumber\\
\dot{\rho}_{43}&=&i\Omega^{\ast}_{c}\rho_{23}-i\Omega_{s}\rho_{42}-[\frac{\Gamma_{1}+\Gamma_{2}}{2}+i(\Delta_{s}-\Delta_{c}+2\Delta_ B)]\rho_{43},\nonumber\\
\dot{\rho}_{22}&=&-(\dot{\rho}_{11}+\dot{\rho}_{33}+ \dot{\rho}_{44}),
\end{eqnarray}
where $\Gamma_1=\gamma_{31}+\gamma_{32}$ and  $\Gamma_2=\gamma_{41}+\gamma_{42}$. The parameter $\gamma_{i1}(\gamma_{i2})(i=3,4)$ denotes the spontaneous decay rate from the excited state $|i\rangle$ to the lower states $|1\rangle$ ($|2\rangle$). The polarization vector in the atomic medium is given by
\begin{equation}\label{e4}
    \vec{P}(z,t)=\chi_{p} \vec{\varepsilon_p} e^{-i(\omega_{p}t-k_{p}z)}+c.c.
\end{equation}
Here, $\chi_{p}$ is the susceptibility representing the response of the medium to the probe field.

Let us now solve the wave equation for the probe field which can be written as

\begin{equation}\label{e5}
\nabla^2{\vec{E}_{p}}-\mu_{0}\epsilon_{0}\frac{\partial^2{\vec{E}_{p}}}{\partial t^2}-\mu_{0}\frac{\partial^2{\vec{P}}}{\partial t^2}=0.
\end{equation}
Inserting equation (\ref{e4}) into equation (\ref{e5}) and using slowly varying approximation, we simplify the equation (\ref{e5}) as
\begin{equation}\label{e6}
\frac{\partial \varepsilon_p}{\partial z}= i2\pi\omega_{p}(\mu_0 \epsilon_0)^{1/2}\varepsilon_p \chi_p.
\end{equation}
Thus the solution of the equation (\ref{e5}) substituting $c=1/\sqrt{\mu_0 \epsilon_0}$ and $k_p=\omega_p/c$  leads the output probe field amplitude to become as
\begin{equation}\label{e7}
\varepsilon_{p}(z=l)=\varepsilon_{p}(0)e^{i2\pi k_p l\chi_p}.
\end{equation}
$\chi_p$ can be related to the probe transition coherence $\rho_{21}$ defined as
 \begin{equation}\label{e8}
\chi_p=\frac{n\mu^2_{21}\rho_{21}}{\hbar \Omega_p},
 \end{equation}
where $n$ is density of atoms and  $\rho_{21}$ is calculated from equation (\ref{e2}). Therefore, the equation (\ref{e7}) reduces to
\begin{equation}\label{e9}
\varepsilon_{p}(z=l)=\varepsilon_p(0)e^{i\frac{\alpha l \rho_{21} \gamma}{2\Omega_p}},
\end{equation}
where $\alpha l=4\pi n\mu^2_{21}kl/\hbar\gamma$ is the resonant absorption. By introducing the normalized susceptibility $S_p=\rho_{21} \gamma/\Omega_p$, the output probe field amplitude takes the form
\begin{equation}\label{e10}
\varepsilon_{p}(z=l)=\varepsilon_{p}(0)e^{i\frac{\alpha l}{2}S_p}.
\end{equation}
 Finally, the normalized transmission of the probe field is given by
\begin{equation}\label{e11}
T=\frac{|\varepsilon_{p}(z=l)|^2}{|\varepsilon_{p}(0)|^2}=e^{-i\alpha l Im[S_p]}.
\end{equation}
The normalized susceptibility $S_p$ is clearly a complex quantity that its imaginary part stands for the absorption of the probe field. The intensity region in which the imaginary part of $S_p$ increases with increase the input intensity denotes the RSA region. Equation (\ref{e11}) displays the transmission behavior of the light which is going to be based on the study of the OL properties of the quantum system.\\
The Z-scan technique is widely used to study the nonlinear refractive index \cite{Sheik} as well as  the OL properties of various materials \cite{Zhou}. In experiment, a Z-scan setup includes a laser field with a transverse Gaussian profile focused by using a lens. The sample is then moved along the propagation direction of the focused Gaussian field. It is clear that the sample experiences maximum intensity at the focal point $(z=0)$, which gradually decreases in either direction from the focus. The Z-scan technique shows the transmission based on the scanning of the sample position relative to the focal plane of the lens. The incident probe field is a Gaussian laser field with the Rabi frequency $\Omega_{p}$
\begin{equation}\label{e12}
\Omega_{p}(z,r,t=0)=\Omega_{p0}\frac{w^{2}_{0}}{w^{2}(z)}exp[-(\frac{2r^{2}}{w^{2}(z)})],
\end{equation}
where $\Omega_{p0}$ is the probe Rabi frequency at the focal point (beam waist), $w_{0}=0.1$ $mm$ is the beam waist radius at focus, $w(z)=w_{0}[1+(z/z_{0})^{2}]^{1/2}$ is the beam radius at $z$ (the distance of the sample from the focal
point) and $z_{0}=\pi w^{2}_{0}/\lambda$ is the diffraction length of the beam. It should be noted that the Z-scan  measurements in our work are carried out for the probe field at $800$ $nm$ wavelength corresponding to the transition $5S_{1/2}\leftrightarrow5P_{1/2}$.\\
With calculations of equation (\ref{e12}) numerically for a pulsed Gaussian beam, normalized transmission as a function of position can be obtained as
 \begin{equation}\label{e13}
T(z)=\frac{4}{w^{2}}\int^{\infty}_{0}$r$ T exp[-(\frac{2r^{2}}{w^{2}(z)})]dr.
\end{equation}

\section*{Results and Discussion}
\begin{figure}[h]
\centering
  \includegraphics[width=8.5cm,height=6cm]{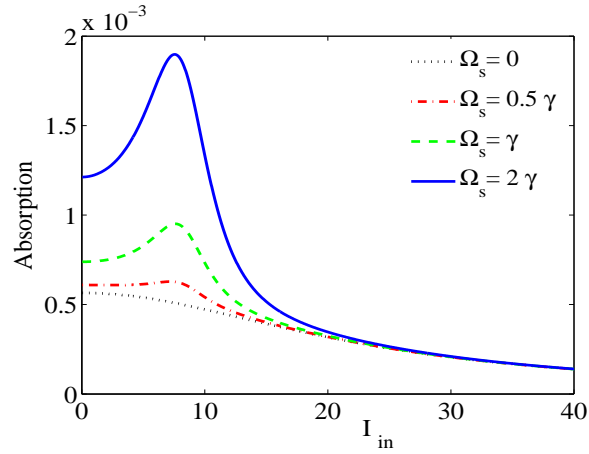}
  \caption{\small Absorption of the probe field versus intensity of the input probe field for different values of $\Omega_s$. The used parameters are $\Omega_{p}=0.01\gamma$, $\Omega_{c}=65\gamma$, $\Delta_c =100\gamma$, $\Delta_{p}=1.5\gamma$, $\Delta_{s}=0$, $\Delta_{B}=0$, $\Omega_s=0$ (dotted), $\Omega_s= 0.5\gamma$ (dot-dashed), $\Omega_s=\gamma$ (dashed) and $\Omega_s= 2\gamma$ (solid).}\label{f2}
\end{figure}

Here, we are going to present our numerical results describing the absorption behavior of the system.  We are interested in investigating the SA and RSA regions in the system to provide the appropriate conditions for inducing the OL. All the parameters are scaled by $\gamma$ which is $2\pi\times 5.75$ MHz for the transition  $5S_{1/2}\leftrightarrow5P_{1/2}$ of $^{87}$Rb atoms. Figure \ref{f2} shows the absorption of the probe field versus the incident intensity of the probe field for different values of $\Omega_s$. The dotted line is for $\Omega_s=0$, the dot-dashed line for $\Omega_s= 0.5\gamma$, the dashed line for $\Omega_s=\gamma$ and the solid line for $\Omega_s= 2\gamma$. The other used parameters are $\Omega_{p}=0.01\gamma$, $\Omega_{c}=65\gamma$, $\Delta_c =100\gamma$, $\Delta_{p}=1.5\gamma$, $\Delta_{s}=0$ and $\Delta_{B}=0$. It is seen that when the coupling field $\Omega_s$ is switched off, the absorption of the probe field decreases by increasing the intensity of the input probe field and the SA is dominant in the absence of the $\Omega_s$.
By switching on the $\Omega_s$, the RSA region  is induced and a peak is generated in the absorption of the probe field, leading to separate the RSA and SA regions. Moreover, the absorption peak enhances by increasing the $\Omega_s$, so it makes us possible to control the RSA phenomenon. Generally, in the RSA region (left side of the peak), the absorption increases by growing the intensity of the input laser field. An investigation on Fig. \ref{f2} shows that the RSA can switch to the SA for the intense input laser field.
\begin{figure}[h]
\centering
  \includegraphics[width=8.5cm,height=6cm]{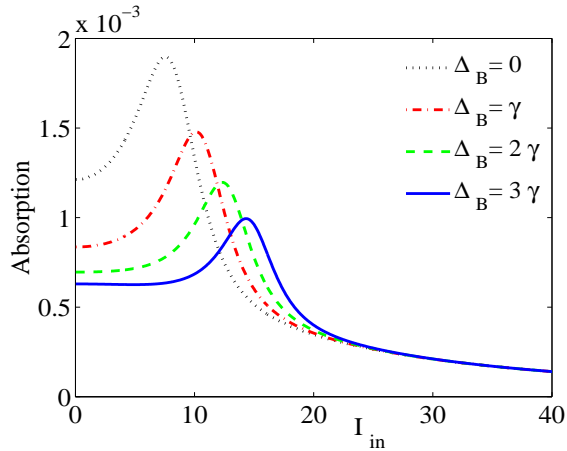}
  \caption{\small Absorption behavior of the probe field versus intensity of the input probe field for different values of $\Delta_B$. The taken parameters are $\Omega_{p}=0.01\gamma$, $\Omega_{c}=65\gamma$, $\Delta_c =100\gamma$, $\Delta_{p}=1.5\gamma$, $\Delta_{s}= 0$, $\Omega_s= 2\gamma$, $\Delta_{B}=0$ (dotted), $\Delta_{B}= \gamma$ (dot-dashed), $\Delta_{B}= 2\gamma$ (dashed) and $\Delta_{B}= 3\gamma$ (solid).}\label{f3}
\end{figure}

The extension of the RSA with respect to the SA region is another scenario that we can do by applying the static magnetic field. After making the RSA in the system, it is important that the system maintains the RSA behavior in wider range of the intensity of the input field. The constructive role of the static magnetic field in the RSA is presented in Fig. \ref{f3} for  $\Delta_{B}=0$ (dotted), $\Delta_{B}= \gamma$ (dot-dashed), $\Delta_{B}= 2\gamma$ (dashed) and $\Delta_{B}= 3\gamma$ (solid). The other taken parameters are $\Omega_{p}=0.01\gamma$, $\Omega_{c}=65\gamma$, $\Delta_c =100\gamma$, $\Delta_{p}=1.5\gamma$, $\Delta_{s}=0$ and $\Omega_{s}=2\gamma$.
We result that the RSA region is extended by increasing the magnitude of the static magnetic field and the RSA is established in a larger range of the input probe field intensity, which promises the extension of the OL range.

\begin{figure}[t]
\centering
  \includegraphics[width=8.5cm,height=6cm]{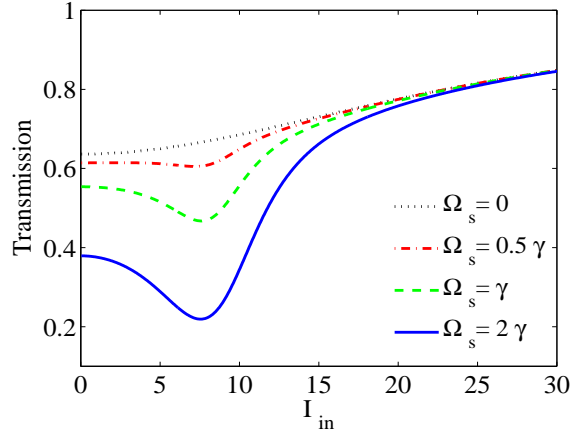}
  \caption{\small Transmission of the probe field versus intensity of the input probe field for $\Omega_s=0$ (dotted), $\Omega_s= 0.5\gamma$ (dot-dashed), $\Omega_s=\gamma$ (dashed) and $\Omega_s= 2\gamma$ (solid). The other used parameters are the same used in Fig. \ref{f2} accompanied by $\alpha l= 800 \gamma$. }\label{f4}
\end{figure}

In the following, we investigate the effect of the coupling field on the transmission of the probe field. In Fig. \ref{f4}, the transmission of the probe field versus the intensity of the input probe field is shown for $\alpha l=800\gamma$, $\Omega_s=0$ (dotted), $\Omega_s= 0.5\gamma$ (dot-dashed), $\Omega_s=\gamma$ (dashed) and $\Omega_s= 2\gamma$ (solid). The other used parameters are those taken in Fig. \ref{f2}. Figure \ref{f4} shows that in the absence of the coupling field, the transmission of the probe field grows with increase of the input probe field intensity going the system toward transparency. Thus, the system cannot be used as  an optical limiter. As proved in Fig. \ref{f2}, the RSA was induced and intensified by increasing the $\Omega_s$. In the RSA domain, the transmission of the probe field keeps constant or even reduces by increasing the coupling field intensity. Hence, it is demonstrated that the OL is coherently induced and controlled in the system. In addition, a bird's eye view of Fig. \ref{f4} reveals that increase of the $\Omega_s$ can lead to decrease the threshold of the OL induced in the system.

\begin{figure}[h]
\centering
  \includegraphics[width=8.5cm,height=6cm]{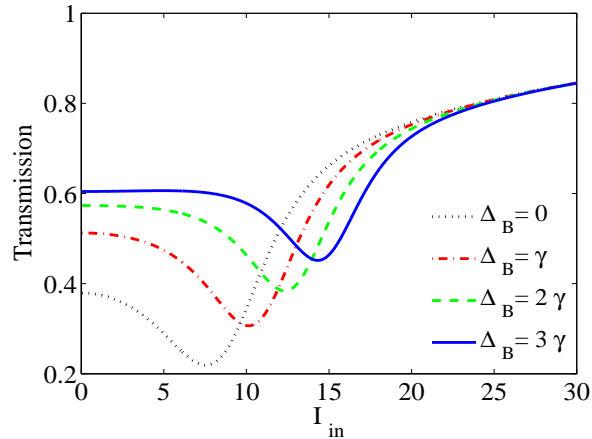}
  \caption{\small Effect of the static magnetic field as $\Delta_{B}=0$ (dotted), $\Delta_{B}= \gamma$ (dot-dashed), $\Delta_{B}= 2\gamma$ (dashed) and $\Delta_{B}= 3\gamma$ (solid) on the transmission of the probe field versus input intensity of the probe field. The other parameters used are the same as those in Fig. \ref{f3} accompanied by $\alpha l= 800 \gamma$. }\label{f5}
\end{figure}

One of the important features that distinguish an optical limiter from the rest is the intensity range in which the optical limiter operates. Here, we show that the applying the static magnetic field can extend the OL range. Figure \ref{f5} depicts the transmission of the probe field versus input intensity for different values of the static magnetic field. The taken parameters are the same used in Fig. \ref{f3}. Figure \ref{f5} shows that the OL range can be controlled by the static magnetic field. It is seen that in the absence of the static magnetic field, the OL is even established in small range of the input intensity. Thus, using the suggested atomic optical limiter, applying the static magnetic field makes the optical devices and sensors safe from damages in a larger range of input intensity.

\begin{figure}[h]
\centering
  \includegraphics[width=8.5cm,height=6cm]{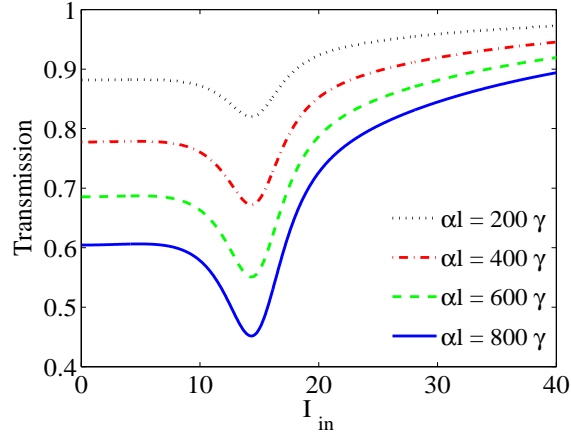}
  \caption{\small Transmission of the probe field versus input intensity of the probe field for different values of the resonance absorption  $\alpha l= 200 \gamma$ (dotted), $\alpha l= 400 \gamma$ (dot-dashed), $\alpha l= 600 \gamma$ (dashed) and $\alpha l= 800 \gamma$ (solid). The other taken parameters are $\Omega_{p}=0.01\gamma$, $\Omega_{c}=65\gamma$, $\Delta_c = 100\gamma$, $\Delta_{p}= 1.5\gamma$, $\Delta_{s}= 0$, $\Omega_s= 2 \gamma$ and $\Delta_{B}= 3 \gamma$. }\label{f6}
\end{figure}

Control of the OL threshold is another advantageous of the suggested optical limiter. In Fig. \ref{f6}, the effect of the resonant absorption, $\alpha l$, is studied on the transmission of the probe field plotted versus the intensity of the input probe field. Resonance absorption is directly related to the length of the medium and density of atoms. It is observed that by increasing $\alpha l$, the OL threshold decreases. Decrease of the OL threshold makes it possible that the presented optical limiter can be set to use in optical devices which needs the optical limiters with lower OL threshold.

\begin{figure}[h]
\centering
  \includegraphics[width=9.5cm,height=6.3cm]{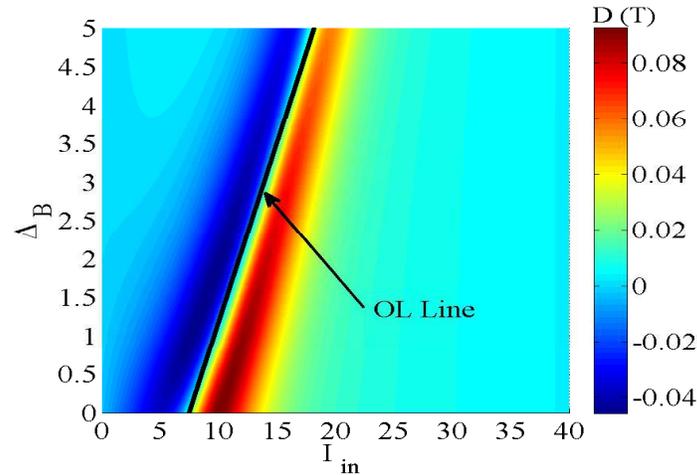}
  \caption{\small Slope of the transmission of the probe field versus the static magnetic field and intensity of the input probe field. The used parameters are $\Omega_{p}=0.01\gamma$, $\Omega_{c}=65\gamma$, $\Delta_c =100\gamma$, $\Delta_{p}=1.5\gamma$, $\Delta_{s}= 0$, $\Omega_s= 2\gamma$ and $\alpha l= 800 \gamma$. }\label{f7}
\end{figure}
In order to gain a deeper insight, the slope of transmission of the probe field, D(T), is displayed in Fig. \ref{f7} as a function of the intensity of the input probe field and the static magnetic field. The used parameters are $\Omega_{p}=0.01\gamma$, $\Omega_{c}=65\gamma$, $\Delta_c =100\gamma$, $\Delta_{p}=1.5\gamma$, $\Delta_{s}= 0$, $\Omega_s= 2\gamma$ and $\alpha l= 800 \gamma$. It is worth to note that for the OL range, the slope of the transmission is zero and even negative. Otherwise, the slope of the transmission is positive for the SA region. Figure \ref{f7} delineates the behavior of the RSA and corresponding the OL induced in the system as well as the SA for all values of the static magnetic field and intensity of the input field. This figure helps us to determine the OL range needed for different optical devices by selecting the appropriate parameters. The OL line, shown in the Fig. \ref{f7}, presents the zero slop of transmission in the end of the RSA region. The left side of the OL line specifies the OL range, while the right side determines the SA region.

\subsection*{Z-scan technique}
\begin{figure}[htbp]
\centering
  \includegraphics[width=8.5cm,height=6cm]{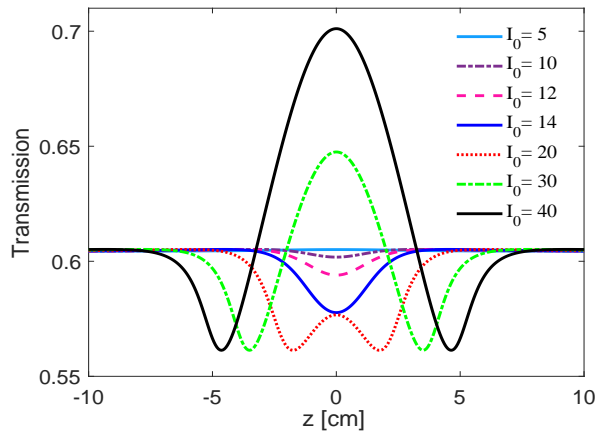}
  \caption{\small Z-scan measurements of the obtained the OL using input gaussian probe field at 800 nm wavelength for different values of input intensity.}\label{f8}
\end{figure}

Finally, we employ Z-scan technique to confirm the validity of the obtained theoretical results. The dip in Z-scan transmission curve corresponds to the OL, while the peak stands for the SA effect. In Fig. \ref{f8}, Z-scan technique measurements is displayed  to investigate the z-dependent transmission for different values of input intensity. As seen in Fig. \ref{f8}, Z-scan technique demonstrates that the OL happens for the input intensities in which the RSA takes place shown in Fig. \ref{f3}. In contrast, for the intensities through the SA region, Z-scan profiles doesn't show the OL around the focal point. The theoretical Z-scan results are in good agreement with the results mentioned in Figs. \ref{f2}-\ref{f6}.

\section*{Conclusion}
In summary, coherent generation and control of the RSA and OL are reported in a four-level Y-type  quantum system. It was shown that the RSA is coherently induced by applied laser fields. We showed that, consequently, the OL is coherently induced through the RSA region so that all characteristics of the induced OL such as the intensity range and the threshold intensity can be controlled by either intensity or frequency of the laser fields. In addition, we proved that the static magnetic field has a constructive role in extending the RSA region and the OL range. Besides, it was demonstrated that the OL threshold can be controlled by either increasing the length of the medium or getting the atomic medium denser. Finally the obtained theoretical results was confirmed by Z-scan technique. Our presented scheme can be used in designing the optical limiters with controllable intensity range and OL threshold.


\section*{Author contributions}
M. Mahmoudi conceived the idea of the research and directed the project. All authors developed the research conceptions, analysed, and discussed the obtained results. M. Ghaderi Goran Abad, M. Mahdieh and M. Veisi performed the calculations. M. Ghaderi Goran Abad wrote the paper with major input from M. Mahmoudi. H. Nadjari has a major role in description of the results.

\section*{Additional information}
\textbf{Competing financial interests:} The authors declare no competing financial and non-financial interests.


\begin{thebibliography}{b}
\bibitem{Zhang} Li, C., Zhang, L., Wang, R., Song, Y. \& Wang, Y. Dynamics of reverse saturable absorption and all-optical switching in $C_{60}$. \textit{J. Opt. Soc. Am. B} \textbf{11}, 1356-1360 (1994).

\bibitem{Tutt}  Tutt, L. W., \&  Boggess, T. F. A review of optical limiting mechanisms and devices using organics, fullerenes, semiconductors and other materials. \textit{Prog. Quantum Electron.} \textbf{17}, 299-338  (1393).

\bibitem{Sun} Sun, Y. P., \& Riggs, J. E. Organic and inorganic optical limiting materials. From fullerenes to nanoparticles. \textit{Int. Rev. Phys. Chem.} \textbf{18}, 43-90 (1999).

\bibitem{Li} Li, Q., Liu, C., Liu, Z., \& Gong, Q.  Broadband optical limiting and two-photon absorption properties of colloidal GaAs nanocrystals. \textit{Opt. Express} \textbf{13}, 1833-1838 (2005).

\bibitem{Ehrlich} Ehrlich, J. E., Wu, X. L., Lee, I. Y., Hu, Z. Y., R\"{o}ckel, H., Marder, S. R., \& Perry, J. W. Two-photon absorption and broadband optical limiting with bis-donor stilbenes. \textit{Opt. Lett.} \textbf{22}, 1843-1845 (1997).

\bibitem{Rao} Rao, S. Optical limiting in the presence of simultaneous one and two photon absorption. \textit{Optik} \textbf{157}, 900-905 (2018).

\bibitem{Venkatram} Venkatram, N., Rao, D. N., \& Akundi, M. A. Nonlinear absorption, scattering and optical limiting studies of CdS nanoparticles. \textit{Opt. Express} \textbf{13}, 867-872 (2005).

\bibitem{Kumar} Venkatram, N., Kumar, R. S. S., \& Narayana Rao, D. Nonlinear absorption and scattering properties of cadmium sulphide nanocrystals with its application as a potential optical limiter. \textit{J. Appl. Phys.} \textbf{100}, 074309 (2006).

\bibitem{Kobyakov} Kobyakov, A., Hagan, D. J., \& Van Stryland, E. W. Analytical approach to dynamics of reverse saturable absorbers. \textit{J. Opt. Soc. Am. B} \textbf{17}, 1884-1893 (2000).

\bibitem{Dong} Dong, P., \& Tang, S. H.  Reverse-saturated absorption and narrow line in the absorption spectrum of a V-type three-level medium. \textit{Phys. Lett. A} \textbf{290}, 255-260 (2001).

\bibitem{Lepkowicz} Lepkowicz, R., Kobyakov, A., Hagan, D. J., \& Van Stryland, E. W. Picosecond optical limiting in reverse saturable absorbers: a theoretical and experimental study. \textit{J. Opt. Soc. Am. B} \textbf{19}, 94-101 (2002).

\bibitem{Azzam} Azzam, S. I., \& Kildishev, A. V. Time-domain dynamics of reverse saturable absorbers with application to plasmon-enhanced optical limiters. \textit{Nanophotonics} \textbf{8}, 145-151 (2018).

\bibitem{Tong} Sun, Z., Tong, M., Zeng, H., Ding, L., Wang, Z., Dai, J., Bian, G., \& Xu, Z. Nanosecond reverse saturable absorption and optical limiting in (Me 4 N) 2 [Cd (dmit)(Sph) 2]. \textit{J. Opt. Soc. Am. B} \textbf{18}, 1464-1468 (2001).

\bibitem{Mishra} Mishra, S. R., Rawat, H. S., \& Mehendale, S. C. Reverse saturable absorption and optical limiting in $C_{60}$ solution in the near-infrared. \textit{Appl. Phys. Lett.} \textbf{71}, 46-48 (1997).

\bibitem{Pan} Pan, H., Chen, W., Feng, Y. P., Ji, W., \& Lin, J. Optical limiting properties of metal nanowires. \textit{Appl. Phys. Lett.} \textbf{88}, 223106 (2006).

\bibitem{Wang} Wang, J., \& Blau, W. J. Inorganic and hybrid nanostructures for optical limiting. \textit{J. Opt.} \textbf{11}, 024001 (2009).

\bibitem{Chen} Liu, Z., Zhang, B., \& Chen, Y. Recent Progress in Two-Dimensional Nanomaterials for Laser Protection. \textit{Chemistry} \textbf{1}, 17-43 (2019).

\bibitem{Muller} Muller, O., Pichot, V., Merlat, L., \& Spitzer, D. (2019). Optical limiting properties of surface functionalized nanodiamonds probed by the Z-scan method. \textit{Sci. Rep.} \textbf{9}, 519 (2019).

\bibitem{Tintu} Tintu, R., Nampoori, V. P. N., Radhakrishnan, P., \& Thomas, S. Reverse saturable absorption in nano colloidal Ge28Sb12Se60 chalcogenide glass. \textit{J. Non-Cryst. Solids} \textbf{375}, 2888-2891 (2011).

\bibitem{Kiran} Kiran, P. P., Reddy, D. R., Maiya, B. G., Dharmadhikari, A. K., Kumar, G. R., \& Desai, N. R. Enhanced optical limiting and nonlinear absorption properties of azoarene-appended phosphorus (V) tetratolylporphyrins. \textit{App. Opt.} \textbf{41}, 7631-7636 (2002).

\bibitem{Lu} Li, F., Zheng, Q., Yang, G., Dai, N., \& Lu, P. Optical limiting properties of two phthalocyanines using 1064-nm laser in solution. \textit{Mater. Lett.} \textbf{62}, 3059-3062 (2008).

\bibitem{Allam} Allam, S. R. Theoretical study on nonlinear properties of four level systems under nano-second illumination. \textit{Laser Phys.}, \textbf{25}, 055701 (2015).

\bibitem{Aziz} Abdullah, M., Bakhtiar, H., Krishnan, G., Aziz, M. S. A., Danial, W. H., \& Islam, S. Transition from saturable absorption to reverse saturable absorption of carmoisine dye under low-powered continuous wave laser excitation. \textit{OPT LASER TECHNOL}, \textbf{115}, 97-103 (2019).

\bibitem{Sheik} Sheik-Bahae, M., Said, A. A., Wei, T. H., Hagan, D. J., \& Van Stryland, E. W. Sensitive measurement of optical nonlinearities using a single beam. IEEE journal of quantum electronics, \textit{IEEE J. Quantum Electron.} \textbf{26}, 760-769 (1990).

\bibitem{Zhou} Liu, Z. B., Liu, Y. L., Zhang, B., Zhou, W. Y., Tian, J. G., Zang, W. P., \& Zhang, C. P. Nonlinear absorption and optical limiting properties of carbon disulfide in a short-wavelength region. \textit{J. Opt. Soc. Am. B} \textbf{24}, 1101-1104 (2007).


\end{thebibliography}
\end{document}